\documentclass[%
 reprint,
showpacs,
 amsmath,amssymb,
 aps,
pre,
]{revtex4-1}

\usepackage{graphicx}
\usepackage{dcolumn}
\usepackage{bm}
\usepackage{subfigure}
\usepackage{color,soul}
\definecolor{lightblue}{rgb}{.80,.85,1}
\sethlcolor{lightblue}

\begin{document}

\preprint{APS/123-QED}

\title{Evolution of a barotropic shear layer into elliptical vortices}

\author{Anirban Guha}
 \email{aguha@mail.ubc.ca}
 \altaffiliation[Also at ]{Institute of Applied Mathematics, University of British Columbia,Vancouver, Canada V6T 1Z2.}
\author{Mona Rahmani}%
\author{Gregory A. Lawrence}%
\affiliation{Department of Civil Engineering, University of British Columbia,Vancouver, Canada V6T 1Z4.}%

\date{\today}

\begin{abstract}
  
When a barotropic shear layer becomes unstable, it produces the well known Kelvin-Helmholtz instability (KH). The non-linear manifestation of KH is usually in the form of spiral billows. However, a piecewise linear shear
layer produces a different type of KH characterized by elliptical vortices of constant vorticity connected via thin braids.~Using direct numerical simulation and contour dynamics, we show that the interaction between two
counter-propagating vorticity waves is solely responsible for this KH formation. We investigate  the oscillation of the vorticity wave amplitude, the rotation and nutation of the elliptical vortex, and straining of the  braids.
Our analysis also provides possible explanation behind the formation and evolution of elliptical vortices appearing in geophysical and astrophysical flows, e.g. meddies, Stratospheric polar vortices, Jovian vortices, Neptune's Great Dark Spot and coherent vortices 
in the wind belts of Uranus.

\end{abstract}


\maketitle

\section{Introduction}

 Barotropic shear layers are ubiquitous in atmospheres and oceans.~These layers can become hydrodynamically unstable, giving rise to an instability mechanism widely
known as the Kelvin-Helmholtz instability (KH). The non-linear manifestation of KH is usually in the form of spiraling billows, whose breaking  generates turbulence and  mixing in
 geophysical flows. 
 
\begin{figure}
\noindent\includegraphics[trim=0cm 0cm 0cm 0cm, clip=true,scale=0.08]{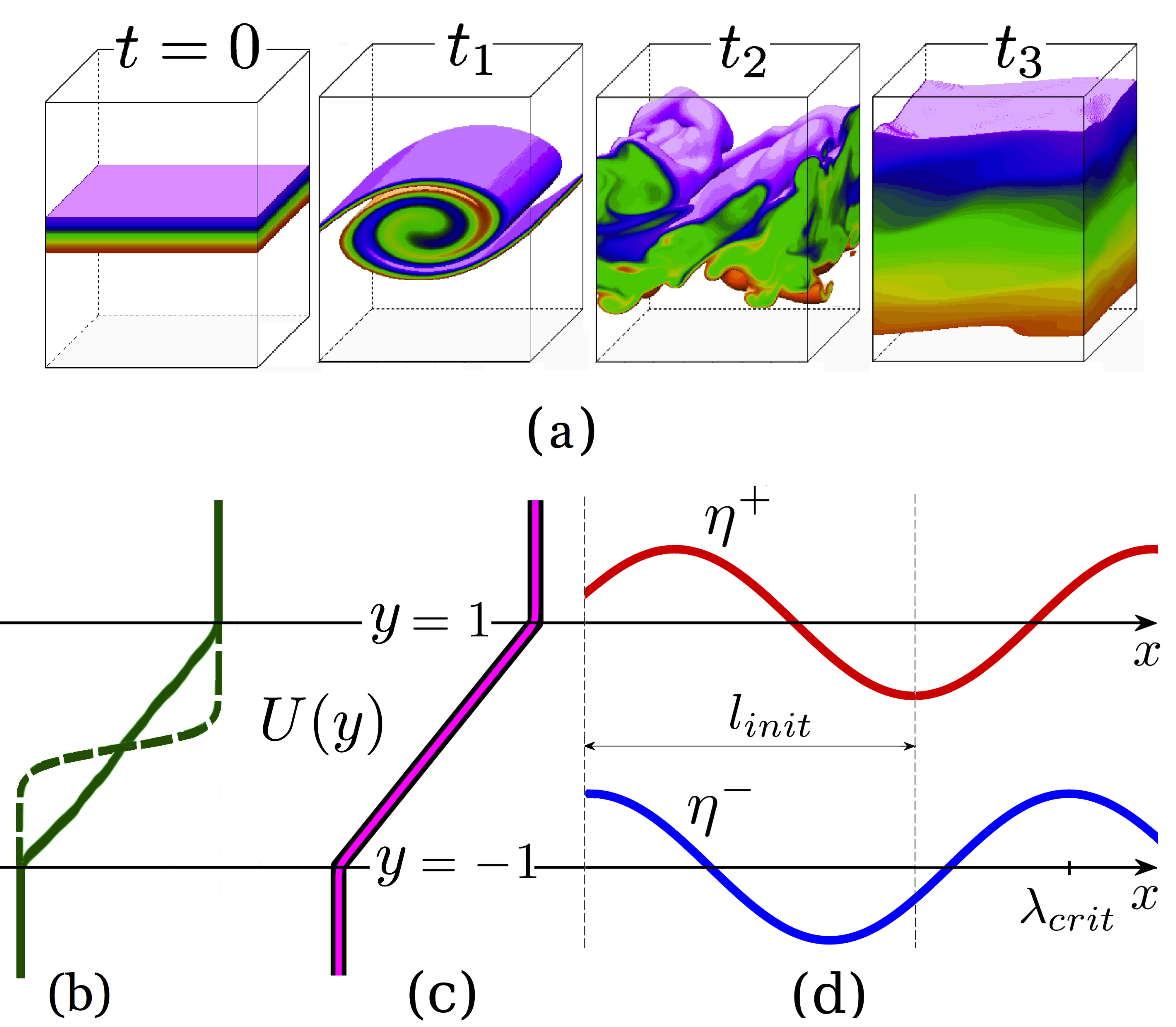}
\caption{(color online) (a) 3D DNS performed to capture the complete turbulent dissipation of a KH billow ensuing from a hyperbolic tangent velocity profile. False color is added to aid visualization. (b) The dashed line represents the initial hyperbolic tangent velocity profile, while the solid line
shows the velocity profile once the flow re-laminarizes ($t=t_{3}$).~(c) The magenta (grey) line is the continuous velocity profile obtained from Eq.\ (\ref{eq:continuousprofile}). This line is drawn on the top of a thick black line, the latter representing the piecewise linear profile from Eq.\ (\ref{eq:1}).  Both the magenta (grey) and black lines closely mimic the solid  line of (b).~(d) Linear vorticity waves (exaggerated) existing at the vorticity discontinuities. 
}
\label{fig:1}
\end{figure}

  In theoretical and numerical studies, the hyperbolic tangent velocity profile is often used to model smooth barotropic shear layers~\cite{haze1972}.~Initially interested in understanding the
  long time evolution of  KH emanating from the hyperbolic tangent profile, we performed a  direct numerical simulation (DNS); see Fig.\ \ref{fig:1}(a).~The flow re-laminarizes once the KH billow completely 
  breaks down into small scales via  turbulent processes.~At this stage, the thickness of the shear layer has approximately quadrupled
  and, more importantly, the profile has almost become piecewise linear (see Fig.\ \ref{fig:1}(b,c)): 
 \begin{equation}
U\left(y\right)=\begin{cases}
1 & y\geq1\\
y & -1\leq y\leq1\\
-1 & y\leq-1\end{cases}\label{eq:1}\end{equation}
Here $U$ is the non-dimensional velocity of the shear flow. Eq.\ (\ref{eq:1}) now becomes the new base flow and serves as the initial condition for the subsequent instability processes. 

The linear stability analysis of the base flow given by Eq.\ (\ref{eq:1}) dates back to Lord Rayleigh~\cite{rayl1880}. The first non-linear  analysis however was performed more than a century later.
Using a boundary integral method
known as \emph{contour dynamics}, Pozrikidis and Higdon \cite{pozr1985} showed that the piecewise linear shear profile evolves into nearly elliptical patches of constant vorticity -  \emph{Kirchhoff vortices}. 
 We  hypothesize that the initial shear
layer profile determines the asymptotic form of the ensuing KH -
smooth shear layers
give rise to spiral billows, while piecewise linear shear layers produce Kirchhoff vortices.

The spiraling billow form of KH has been thoroughly investigated in the past.  In fact, the spiral billow
shape has  become the signature of KH  \cite{smyth2012};
but little is known about the non-linear evolution of the piecewise linear shear layer. 
This is because the  piecewise linear profile is usually considered to be of little practical relevance, hence its usage is  restricted to 
theoretical studies - mainly as an approximation to smooth shear layers \cite{pozr1997,carp2012}.~On the contrary, our DNS result in Fig.\ \ref{fig:1}(a,b) indicates that the  quasi piecewise linear  profile is also likely to 
occur in nature. The fact that this profile produces elliptical vortices 
similar to those observed in geophysical and astrophysical flows,
e.g.\ meddies in Atlantic ocean  \cite{Cher2003},  stratospheric polar vortices \cite{waugh1999}, Great Red Spot and other Jovian vortices \cite{morales2002}, Neptune's Great Dark Spot \cite{poli90}, and coherent vortices in the atmosphere of Uranus \cite{liu2010}, has motivated us to investigate further. 

\section{Linear theory}

In 1880,  Lord Rayleigh \cite{rayl1880}  performed a linear stability analysis of the shear layer profile in Eq.\ (\ref{eq:1}) and showed it to be
 unstable for the range of wavenumbers $0\leq\alpha\leq0.64$, 
 the fastest growing mode being $\alpha_{crit}=0.4$. In the conventional linear stability approach, infinitesimal wavelike perturbations are superimposed on a laminar background flow and 
 an eigenvalue problem is solved to find  the band of unstable wavenumbers  \cite{draz1982}.~This mathematical exercise, however, provides little insight
 into the underlying physical mechanism.~In the past $50$ years there has been a continuous effort to provide a mechanistic picture of hydrodynamic instabilities, especially homogeneous and stratified shear instabilities. Analytical studies performed on the profile in Eq.\ (\ref{eq:1}) have shown that the interaction between two vorticity waves is responsible for the development of KH \cite{holm1962,bret1966,cair1979,hosk1985,caul1994,bain1994,carp2012,guha_wave_interaction2012}. Following Guha and Lawrence \cite{guha_wave_interaction2012}, we will refer to this 
 wave interaction based interpretation of hydrodynamic stability theory as  ``wave interaction theory'' (WIT).

 KH can be understood from the perspective of WIT  by referring to  Eq.\ (\ref{eq:1}). The vorticity $\Omega \equiv dU/dy$
is discontinuous at $y=\pm1$, which allows each of these two locations to support
 a \emph{stable}, progressive, interfacial wave called
the vorticity wave (also known as the Rayleigh wave). In a rotating frame,
its analogue is the Rossby edge wave which exists at the discontinuities in
potential vorticity.~In general,  a  vorticity wave at the interface $y=y_{j}$
 has a phase speed $c_{r}(y_{j})$  given by 
 \begin{equation}
 c_{r}(y_{j})= U(y_{j})+c_{r}^{int}(y_{j}) 
 \label{eq:wave}\end{equation}
The first component, $U(y_{j})$, is the background velocity, while $c_{r}^{int}(y_{j})$ is the intrinsic phase speed defined as
 \begin{equation}
 c_{r}^{int}(y_{j})= \frac{\left[\Omega\right]_{y_{j}}}{2\alpha}
 \end{equation}
where $\left[\Omega\right]_{y_{j}}=\Omega(y_{j}+\epsilon)-\Omega(y_{j}-\epsilon)$
is the jump in $\Omega$ across $y_{j}$. For the profile in Eq.\ (\ref{eq:1}), the arrangement is such that $c_{r}^{int}(y_{j})$ is always directed counter to $U(y_{j})$. We define such waves as  counter-propagating
vorticity waves (CVWs) (analogous to counter-propagating Rossby waves
in a rotating frame). The CVWs  at $y=\pm1$ have  \cite{caul1994}: 
 \begin{equation}
c_{r}^{\pm}\equiv c_{r}\left(\pm1\right)=\pm1\mp\frac{1}{2\alpha}
 \end{equation}
 The above equation implies that these two waves always travel in opposite directions. The wave at $y=1$ is left moving while the one at $y=-1$ moves to the right.
Although each of them is marginally stable, their interaction can lead to instability, producing KH.

Consider a pair of CVWs:
\begin{eqnarray}
 & & \eta^{+}=a_{0}\cos\left(\alpha x-\Phi\right) \,\,\,\,\,\,\,\,\textrm{at}\,\,y=1 \label{eq:topwave}\\
 & & \eta^{-}=a_{0}\cos\left(\alpha x\right) \,\,\,\,\,\,\,\,\,\,\,\,\,\,\,\,\,\,\,\,\textrm{at}\,\,y=-1 \label{eq:bottomwave}\end{eqnarray}
where the perturbation amplitude $a_{0}\ll1$ and the phase shift $\Phi\in\left[-\pi,\pi\right]$. \emph{Modal growth} occurs only when the two waves together behave like a normal mode. Two conditions  need to be precisely satisfied for modal growth:
 (a) \emph{phase locking} - the waves
are stationary relative to each other, i.e.\ $c_{r}^{+}$ and $c_{r}^{-}$ become zero after interaction, and (b) \emph{mutual growth} - the
phase shift between the waves is such that one makes the other grow exponentially. Linear theory predicts this phase shift to be $\Phi^{modal}=\cos^{-1}\left\{ \left(1-2\alpha\right)e^{2\alpha}\right\}$  \cite{guha_wave_interaction2012}. For $\alpha=\alpha_{crit}$,
the corresponding  $\Phi_{crit}^{modal}=0.353\pi$.
Thus WIT provides the appropriate initial conditions (i.e.\ $\alpha_{crit}$ and $\Phi_{crit}^{modal}$) for modeling the non-linear evolution. 

Unfortunately, WIT itself is limited to the \emph{linear} regime only. Current 
methodology does not allow a straightforward non-linear extension. This is a direct consequence of linearization which forces the vorticity interfaces at $y=y_{j}$ to become vortex sheets \cite{holm1962,bain1994,carp2012}.
Therefore, it remains to be shown
 whether the \emph{non-linear} evolution of KH can be understood in terms of two interacting non-linear vorticity waves.

\section{Non-linear formulation}

 Over the last few decades, sophisticated computational techniques have been developed for precise understanding of turbulent processes.
 Two such techniques worth mentioning are direct numerical simulation (DNS) and vortex methods. 
 
 Many geophysical and astrophysical flows can be assumed homogeneous, incompressible
 and quasi-inviscid. In such flows, vorticity plays a major role in driving non-linear processes like chaos and turbulence \cite{saff1995}.
 Vortex methods are especially useful under such circumstances; they numerically solve the inviscid and incompressible Navier-Stokes equations (Euler equations). An example of one such 2D vortex method is contour dynamics \cite{deem1978}.

\subsection{Contour dynamics}

High Reynolds number flows have a tendency to develop finite-area vortex regions or ``vortex patches'' with steep sides  \cite{deem1978,pull1992}. If the flow is  incompressible and 2D,  the area as well as the vorticity of a vortex patch  are conserved quantities. The vortex patch boundary is referred to as a ``contour'', and is both a vorticity jump
and a material surface \cite{saff1995}. A small perturbation on this contour sets up a vorticity wave \cite{deem1978}.  

  Substantial simplification 
is possible for constant vorticity patches; the governing 2D Euler equations can be
reduced to a 1D boundary integral.~This  provides a significant  advantage for computing the vortex patch evolution
by only solving the contour motion. This methodology is known as  contour dynamics (CD) \cite{pull1992}.~Numerical implementation of CD is done in a Lagrangian framework by tracing the
contour with a set of $N$ marker points. 

Observing that a piecewise linear shear layer can be represented by a horizontally periodic patch of constant vorticity, 
Pozrikidis and Higdon \cite{pozr1985}  have used CD  to simulate the non-linear evolution process. The evolution of the $i$-th Lagrangian
marker is given by \cite{pozr1997} :
\begin{equation}
\frac{d\mathbf{x}_{i}}{dt}=-\frac{\Omega}{4\pi}\intop_{C}\ln\left[\cosh\left(\alpha\triangle y_{i}^{'}\right)-\cos\left(\alpha\triangle x_{i}^{'}\right)\right]d\mathbf{x^{'}}
\label{eq:evolution-1} 
\end{equation}
where $\Omega=1$,  $\mathbf{x}=\left[x,y\right]^{T}$, $\triangle x_{i}^{'}=x_{i}-x^{'}$, $\triangle y_{i}^{'}=y_{i}-y^{'}$ and $C$ is the contour around
one patch. On perturbing the contour with sinusoidal disturbances
, the shear layer rolls up
 producing nearly elliptical patches of constant vorticity  \cite{pozr1985}. 
 
\emph{ We interpret the rolling up of the piecewise linear shear layer in terms of the interaction between two CVWs.}
To better understand this point, let us first consider the simple case of a circular patch of constant vorticity. A small perturbation on its contour produces a linearly stable vorticity wave (also known as a ``Kelvin wave'') \cite{deem1978,mitc2008}. Similarly, perturbation on a periodic, piecewise linear shear layer produces  two counter-propagating vorticity waves, one at the upper interface and the other at the lower interface.  These two CVWs interact with each other causing the shear layer to evolve into elliptical vortex patches. Since WIT represents the linear dynamics of the shear layer evolution process, CD can be regarded as a non-linear extension of WIT. 

\begin{figure*}
\noindent\includegraphics[trim=0cm 0cm 0cm 0cm, clip=true, scale=0.84]{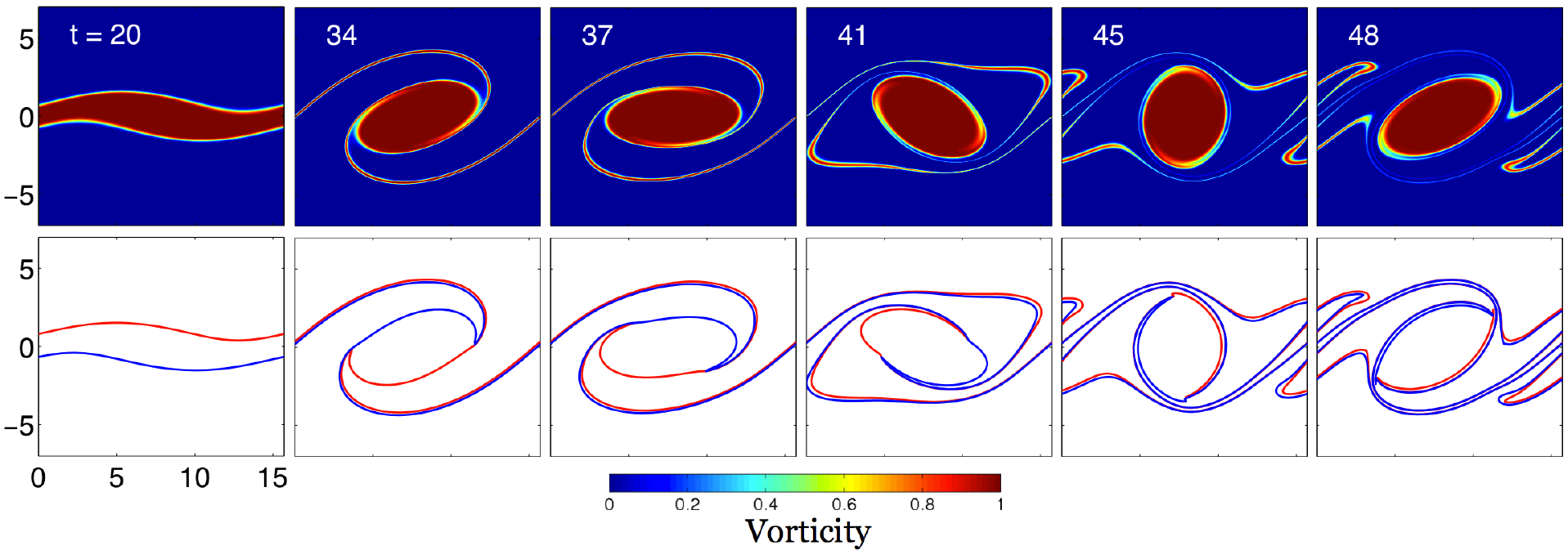}
\caption{(color online) Time evolution of Kelvin-Helmholtz instability - comparison between DNS (top) and CD (bottom).}
\label{fig:2}
\end{figure*}

In the present study we follow the approach of Pozrikidis and Higdon \cite{pozr1985} except for our choice of $\alpha$ and $\Phi$ in Eqs (\ref{eq:topwave})-(\ref{eq:bottomwave}).  We set $\alpha= \alpha_{crit}$ to ensure that the resulting non-linear structure is most likely to be realized in nature; whereas,  Pozrikidis and Higdon used several values of $\alpha$, not including $\alpha_{crit}$.  We also choose $\Phi=\Phi_{crit}^{modal}=0.353\pi$ to ensure exponential growth from time t = 0; whereas, in  Pozrikidis and Higdon, $\Phi = 0$ or $\pi$.
Considering a domain one wavelength ($\lambda_{crit}\equiv 2\pi/\alpha_{crit}=5\pi$) long, we solve Eq.\ (\ref{eq:evolution-1})~using central differencing for 
space derivatives and $4^{\textrm{th}}$ order Runge-Kutta for time.
Each wave is initially represented by 400 points.~During its evolution process, an adaptive point insertion-deletion algorithm is used to check if the neighboring points are within a desired distance.

\subsection{Direct numerical simulation}
Previous studies have  used CD mainly as a tool for \emph{qualitative} understanding of problems involving inviscid  vortical flows \cite{pull1992}.
In order to demonstrate the quantitative capabilities of CD,
 we validate our CD simulation against a pseudo-spectral DNS. 
 The DNS code uses full Fourier transforms in the horizontal direction, and 
 half-range sine or cosine Fourier transforms in the vertical direction in order to convert the set of partial differential equations (Navier-Stokes equations) into ordinary differential equations. Time integration is 
 performed using a third-order Adams-Bashforth method. Detailed description of this code can be found in Winters et al.\ \cite{wint2004}.

  We consider a domain of 
 length $\lambda_{crit}$ in the horizontal direction and nine times the initial shear layer thickness in the vertical direction.  The horizontal boundary condition is periodic while the vertical boundary condition is no-flux free-slip. 
 We perform  a 2D simulation at  Reynolds number  $Re=10,000$ ($Re=1/\nu$, where $\nu$ is the fluid viscosity). 
This Reynolds number is high enough to mimic quasi-inviscid flow conditions. To simulate the smallest scales of motion, 
 we resolve our domain using $2880$ points in horizontal and $3456$ points in vertical direction.
 
Since non-differentiable profiles like Eq.\ (\ref{eq:1}) are subject to Gibbs phenomena,
we use a smooth velocity profile that resembles the piecewise linear profile
very closely; see Fig.\ \ref{fig:1}(c). This velocity profile is derived from a vorticity distribution having the form
\begin{equation}
\Omega\left(y\right)=\frac{1}{2}\left[1-\tanh\left(\frac{y^{2}-1}{\epsilon}\right)\right]\label{eq:continuousprofile}\end{equation}
Integrating Eq.\ (\ref{eq:continuousprofile}), the velocity profile is obtained directly: $U=\int\Omega dy$. The linear stability characteristic of this profile matches  with 
that of the piecewise linear profile almost exactly, and has the same $\alpha_{crit}$. Equating the total circulation of DNS with CD yields $\epsilon=0.100$.  The vorticity field in DNS is perturbed to match the initial wave amplitude growth in CD.

\begin{figure}
\includegraphics[trim=0cm 0cm 0cm 0cm, clip=true, scale=0.22]{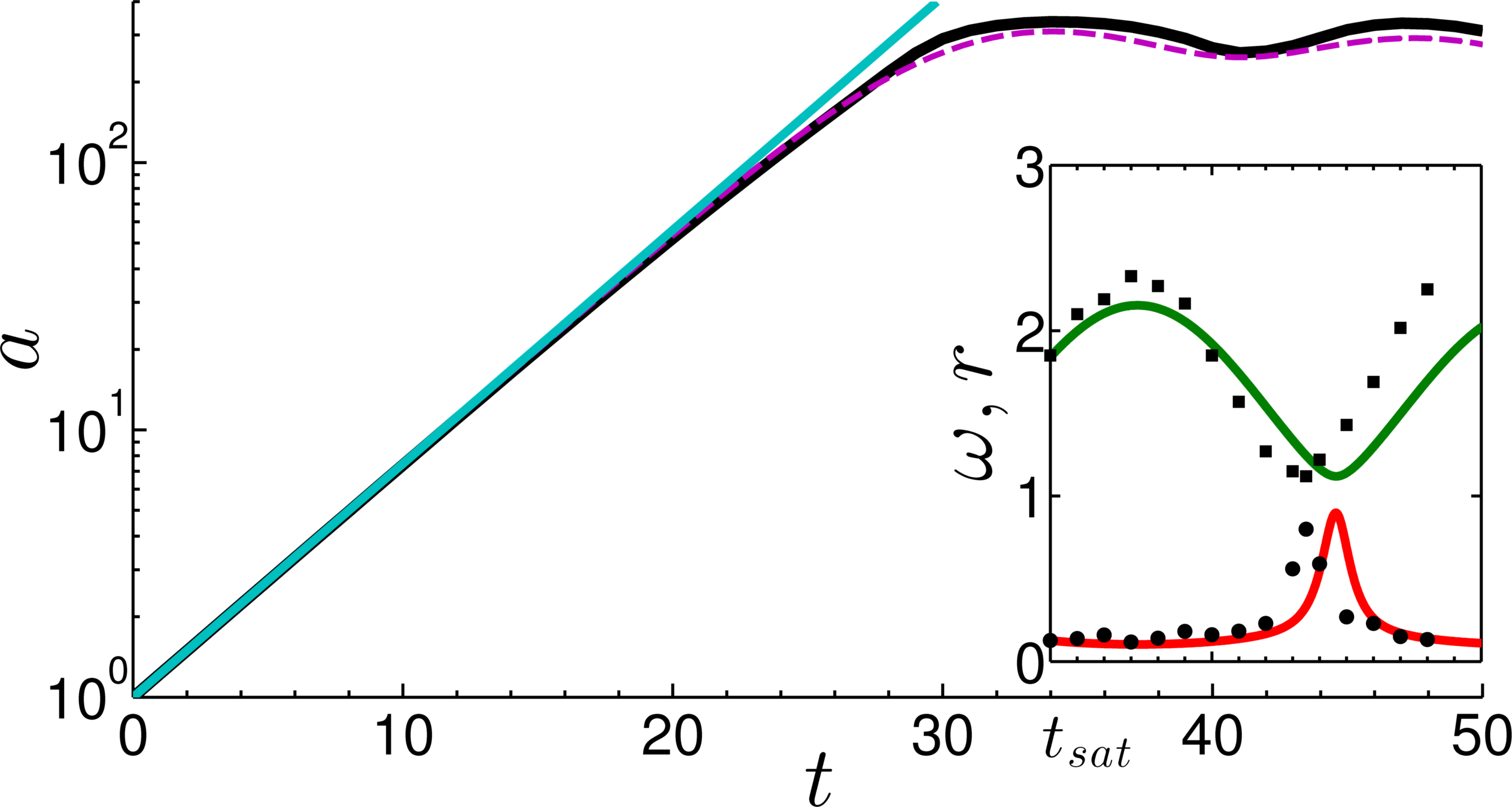}
\caption{(color online) Temporal variation of the wave amplitude $a$. The straight cyan (light grey) line is the prediction from linear theory, the black line corresponds to CD while the dashed magenta (dark grey) line represents DNS results.
 The green (top) and red (bottom) lines in the inset respectively show the variation of the ellipse aspect ratio $r$ and the angular rotation rate $\omega$ with time. These variations are obtained by solving Eqs.\ (\ref{eq:thetadot})-(\ref{eq:rhodot}). 
 The black markers indicate the corresponding data points measured from DNS.}
\label{fig:3}
\end{figure}

\section{Results and discussion}

\subsection{Pre-saturation and saturation phases}
The non-linear evolution of KH is illustrated in Fig.\ \ref{fig:2}. It shows the vorticity field from DNS and contour lines from CD. The vorticity of the region enclosed by the contour lines is conserved in CD.
However the presence of viscosity makes conservation of vorticity invalid in DNS. The implementation of high $Re$ DNS minimizes the viscous effects, making DNS comparable to CD. 
The basic premise behind our simulations is the conservation of total circulation $\Gamma=\Omega A$ (where the vorticity $\Omega=1$ and $A$ is the shear layer area), which comes from Kelvin's Circulation theorem \cite{saff1995}. Its corollary  is the conservation of shear layer area -
a quantity that remains fixed at its initial value $A=2\lambda_{crit}$. 

Fig.\ \ref{fig:3} shows the time evolution of the wave amplitude $a=|\max(\eta^{+})-1|=|\min(\eta^{-})+1|$. The maximum shear layer thickness, $H=2(1+a)$ evolves in a fashion similar to the wave amplitude $a$.
 We find the growth to be exponential, at least for $t\lessapprox20$. CVW interaction causes  the shear layer to grow non-linearly. This phenomenon leads to the roll-up and formation of the elliptical core vortex (Kirchhoff vortex).  
 The evolution process is shown in Fig.\ \ref{fig:2}. The part of the shear layer  between the crest of the lower wave and the trough of the upper wave (see Fig.\ \ref{fig:1}(d)) gives rise to the elliptical core, the initial length of which is given by
 \begin{equation}
  l_{init}=\left(1+\frac{\Phi_{crit}^{modal}}{\pi}\right)\frac{\lambda_{crit}}{2}
 \end{equation}
 
 The flow saturates  (i.e. the amplitude reaches a maximum) at $t_{sat}=34$. For $t \ge t_{sat}$, approximately $80\%$ of $\Gamma$ is concentrated in the core. H  also reaches a maxima at saturation, and has the value $H_{max}=8.7$. The fully formed elliptical cores  are connected by thin filaments of fluid called \emph{braids}. 
 These braids wind around the rotating cores, see Fig.\ \ref{fig:2}.

\subsection{Early post-saturation phase}
After saturation, the core  rotates with an angular velocity  $\omega$, causing the wave amplitude, a, to oscillate with a time period $T_{amp} \approx 13$; see Fig.\ \ref{fig:3}.  
The core also nutates, i.e.\ the core aspect ratio $r$ (defined as the ratio between the ellipse major axis and  the minor axis) undergoes a periodic oscillation. This \emph{nutation} phenomenon is  apparent in both Figs.\ \ref{fig:2} and \ref{fig:3}. 
\subsubsection{Nutation}
To better understand the nutation process, 
we consider the simple model proposed by Kida \cite{kida1981}. An isolated Kirchhoff vortex rotates in the presence of a constant background strain-rate $\gamma$. 
The velocity field associated with this strain-rate is given by
$u_{s}=\gamma\sigma$, $ w_{s}=-\gamma\xi$
where $\sigma$ and $\xi$ are the principal axes with the origin at the centre of the ellipse. In our case, this  velocity field mimics the leading order straining effect induced by the rotation of
other Kirchhoff vortices. Note that the periodic boundary condition takes into account the effects of other Kirchhoff vortices. 

Let  the clockwise angle between $\sigma$ and the ellipse major axis be $\theta$ at any instant.~Then $\theta$ and $r$ evolve as follows \cite{kida1981}:
\begin{eqnarray}
 &  & \omega \equiv \frac{d\theta}{dt} =-\gamma\left(\frac{r^{2}+1}{r^{2}-1}\right)\sin\left(2\theta\right)+\frac{\Omega r}{\left(r+1\right)^{2}}\label{eq:thetadot}\\
 &  & \frac{dr}{dt}=2\gamma r\cos\left(2\theta\right)\label{eq:rhodot}\end{eqnarray}
Eq.\ (\ref{eq:rhodot})  implies that the nutation is caused by strain.~It also reveals that $r$ reaches maxima at $\theta=\pm \pi/4$. Simultaneously, Fig.\ \ref{fig:2} shows that  the core nutates with a maximum value of
$r$  along the $x$ axis and a minimum along the $y$ axis. Therefore the $\sigma$ axis must make an angle of $\pi/4$ with the $x$ axis. The  angle made by the braid with the $x$ axis at the stagnation point(s) is also $\pi/4$.~This is because the
 braid aligns itself with the streamlines.
 
 To investigate why the  $\sigma$ axis makes an angle of $\pi/4$ with the $x$ axis, we consider an idealized problem where an infinite number of  Kirchhoff vortices, each of
 circulation $\Gamma_{core}=\Gamma$ (note $\Gamma=2\lambda_{crit}$), are placed along the $x$ axis with a constant spacing  $\lambda_{crit}$ between their centers. 
Furthermore we replace all the Kirchhoff vortices with point vortices of the same strength, i.e.\ $\Gamma_{core}$. This provides a simplistic understanding of the mechanism by which the rotation of distant vortex patches strain a given patch. 
We find this \emph{ideal} strain-rate to be
\begin{equation}
\gamma'=\frac{\Gamma_{core}}{2\pi\lambda_{crit}^{2}}\sum_{n=-\infty, n\neq0}^{\infty}n^{-2}=0.067
\label{eq:idealvort}\end{equation}
The principal axes of the strain field produced by this infinite array of point vortices make angles of $\pm \pi/4$ with the $x$ axis.
Hence, this ideal strain field and the strain field of our actual problem have the same orientation. Before comparing the magnitudes of these two fields, it is important to note  that the  presence of  braids complicate the actual problem by making the strain-rate magnitude vary spatially. We simplify the analysis by  assuming a  strain field of constant magnitude acting on the elliptical core, thereby reducing the problem to the Kida problem described by
  Eqs.\ (\ref{eq:thetadot})-(\ref{eq:rhodot}). DNS is used to supplement the analysis by providing the values of $r$ and $\theta$ wherever necessary. By applying this methodology, the 
 magnitude of the actual strain-rate is found to be $\gamma=0.073$, which is  close to the ideal value of $0.067$ obtained from Eq.\ (\ref{eq:idealvort}).  

We also capture the evolution of $r$ and $\omega$ by solving Eqs.\ (\ref{eq:thetadot})-(\ref{eq:rhodot}); see the inset in Fig.\ \ref{fig:3}. The initial values are obtained from DNS, and $\gamma=0.073$. 
The spike in $\omega$ at $t\approx 45$ is caused by $r \rightarrow 1$.  The figure shows that Kida's model compares well with the DNS. 
The nutation period is found to be $T_{nut}\approx 13$, while the period of core rotation is $T_{core}=2\pi/\bar{\omega}\approx 26$ 
(overbar denotes average). $T_{core} \approx 2 T_{nut}$ is because one full rotation corresponds to passing the coordinate axes twice. Likewise, $T_{core} \approx 2 T_{amp}$ because  the braids are connected to the two ends of the core.

\begin{figure}
\includegraphics[trim=0cm 0cm 0cm 0cm, clip=true, scale=0.45]{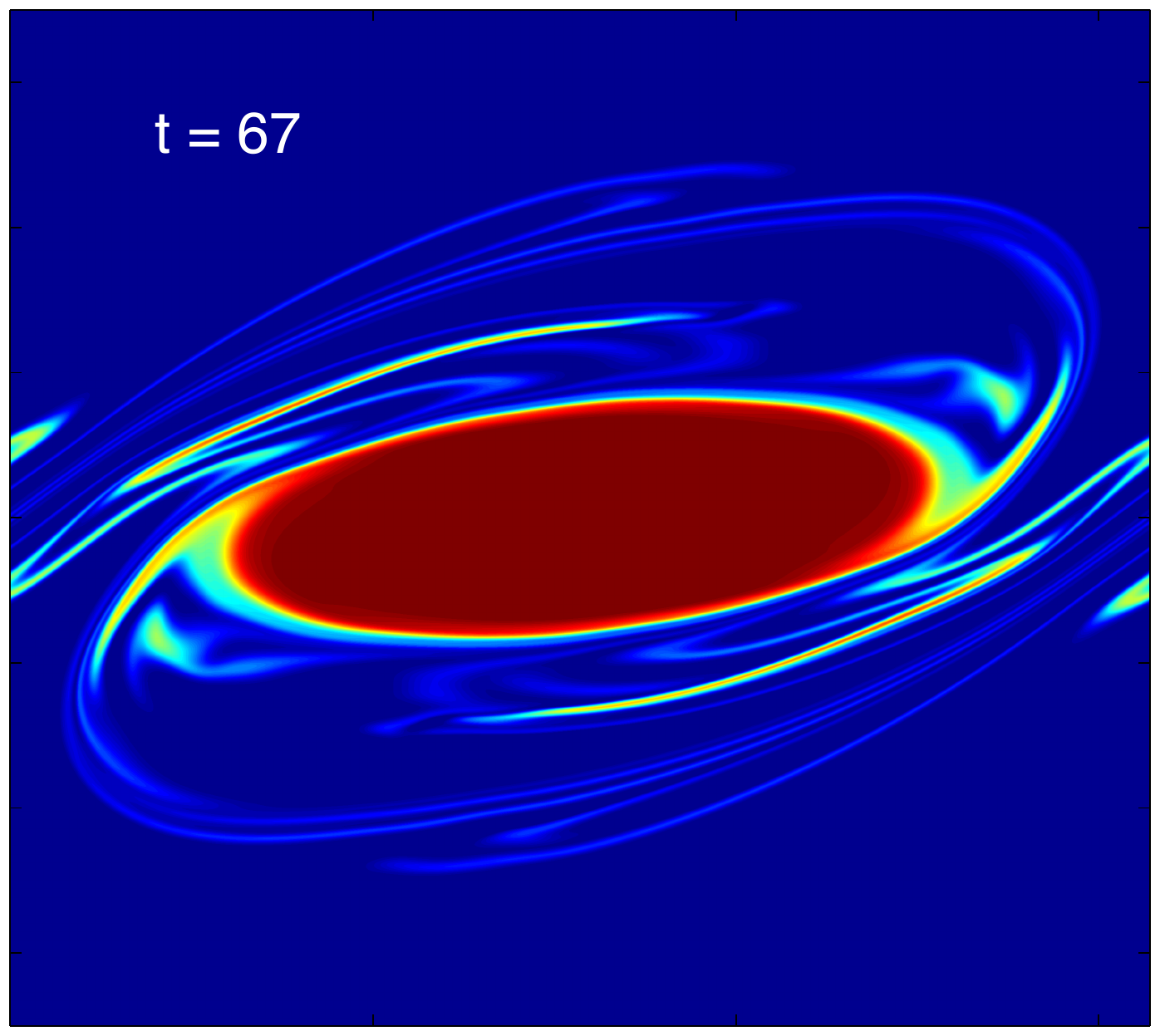}
\caption{(color online) Formation of winding filaments around the elliptical vortex during the late post-saturation phase.}
\label{fig:4}
\end{figure}

\subsubsection{Small length scale production}
The smallest length scales are found to occur in the braid region adjacent to the core; see Fig.\ \ref{fig:2}. This is due to the straining effect of the rotating elliptical core which causes the braid region in its vicinity to  thin exponentially fast. In real flows, 
when a fluid element becomes sufficiently thin, the balance between the strain-rate and the viscous dissipation determines the small length scales. The order of magnitude of the  core rotation induced  strain-rate, $\gamma_{local}$,  can be obtained by replacing the core with a point vortex of equivalent
 strength and located at the ellipse centre: 
 \begin{equation}
 \gamma_{local}\sim \frac{\Gamma_{core}}{2\pi l^{2}}=\frac{1}{2}
 \end{equation}
 where $l$ is the characteristic length of the core. Notice that this local strain-rate is one order of magnitude greater than the background strain-rate $\gamma$ or $\gamma'$.
 
The smallest length scale appearing in a 2D ``turbulent'' flow is  $L_{2D} \sim Re^{-1/2}$ \cite{david2004}. This length scale is a 2D analogue of the Taylor microscale occurring in 3D turbulent flows. 
In order to estimate the time when $L_{2D}$  appears in our flow, we formulate a braid evolution equation similar to Eq.\ (2.8) of Corcos and Sherman \cite{Corc1976}:
\begin{equation}
\delta^{2}\left(t^{*}\right)=\delta^{2}\left(0\right)e^{-2\gamma_{loc} t^{*}}+\frac{\pi}{2\gamma_{loc} Re}\left( 1-e^{-2\gamma_{loc} t^{*}}\right)
\label{eq:thinning}\end{equation}
where $\delta\left(t^{*}\right)$ is the braid thickness  adjacent to the core at time $t^{*}=t-t_{sat}$.  We estimate $\delta\left(0\right)$ from our DNS and solve Eq.\ (\ref{eq:thinning}). $L_{2D}$ is found to
appear soon after saturation, around $t^{*}\approx 4$, implying that 2D transitional flows like KH can give rise to ``turbulent'' features at a very early stage.

\subsection{Late post-saturation phase}

For $t\gtrapprox50$, the core surface develops progressive vorticity waves. These are called the Kelvin $m$-waves or the Love $m$-waves \cite{mitc2008}, where $m$ is the eigenmode. In the absence of strain, $r<3$ is 
the condition for stability of a Kirchhoff vortex  \cite{saff1995}. Although $r$ satisfies this condition in our case (refer to the inset in Fig.\ \ref{fig:3}), the presence of strain adds instability \cite{Drit1990}. Finite amplitude $m=4$ waves  give rise to winding filaments, see Fig.\ \ref{fig:4}. This is a  feature of Kirchhoff vortices in background shear or strain, which has been thoroughly investigated by Dritschel \cite{Drit1990}. Background shear or strain is however not necessary to produce winding filaments.  Similar winding features are also seen in  large aspect ratio Kirchhoff vortices  even in the absence of shear or strain \cite{mitc2008}.
 
 \section{Practical implications}
Elliptical  cores, similar to those shown in our simulations, are a common feature in 
geophysical and astrophysical flows, e.g.\ Mediterranean eddies or meddies  \cite{Cher2003},   
 vortices in Jupiter  \cite{poli90,morales2002,liu2010}, the Great Dark Spot of Neptune  (which has disappeared now) \cite{poli90}, coherent vortices in the wind belts of Uranus \cite{liu2010} and stratospheric polar vortices \cite{waugh1999}.

Elliptical vortices  known as  \emph{meddies} are found in the region where Mediterranean sea meets the Atlantic ocean. The genesis of meddies is highly speculative; one
 possible explanation is the interaction between counter-propagating Rossby waves. This is corroborated by the observations
 near Portim\~ao Canyon and Cape St.\ Vincent  \cite{Cher2003}. 
 The average diameter of a meddy is approximately half of the most unstable wavelength of the Rossby edge waves  \cite{Cher2003}. 
 Indeed, our simulations reveal the same ratio between the average diameter of the elliptical core and $\lambda_{crit}$. Hence we conjecture that  meddies
 may be  a non-linear manifestation  of KH ensuing from quasi piecewise linear shear layers.  Moreover, our DNS also captures the winding filaments observed during the meddy evolution process \cite{mene2012}, see Fig.\ \ref{fig:4}. 
 
 The Great Red Spot and White Ovals of Jupiter, and the Great Dark Spot of Neptune can be well  
approximated as Kirchhoff vortices \cite{poli90}. Astronomical observations  reveal that these 
vortices undergo rotation and nutation, which can be well predicted using the Kida model \cite{poli90}. Elliptical vortices are also found in the wind belts of Uranus, see Fig.\ 3 of Liu and Schneider \cite{liu2010}. Moreover  earth's  stratospheric polar vortices are also nearly elliptical in shape \cite{waugh1999}. 
Hence we hypothesize that  planetary vortices, just like meddies,  may have been produced by the interaction between counter-propagating Rossby waves.

It is worth mentioning that meddies and planetary vortices are almost always embedded in a background shear flow. Hence proper modeling of  these vortices require an initial background velocity profile different from  (quasi) piecewise linear.~Background shear, much like background strain,  causes angular acceleration and nutation of the elliptical vortices. Background shear can be easily incorporated in CD, DNS and the Kida model,  making the subsequent analysis a straightforward extension of our present study.

Geophysical and astrophysical flows are usually density stratified, hence vorticity can be generated due to baroclinic effects. Moreover, three dimensionality can become important with the emergence of small scale features like braids and secondary structures on the vortex boundary.  Furthermore, vortices might merge leading to an inverse cascading of energy. Simulating vortex merging requires considering multiple wavelengths. All the above-mentioned factors might be essential for a detailed understanding of geophysical and astrophysical vortices, but their consideration will  complicate the problem and increase the computational cost. 

\section{Conclusion}
When a piecewise linear shear layer becomes unstable, it  evolves into a series of elliptical vortices of constant vorticity (Kirchhoff vortices) connected by thin braids. The interaction between two counter-propagating vorticity waves is the driving mechanism behind this instability process.~Although this wave interaction perspective was known previously, linearized approximations forced the
analysis to be valid only in the linear regime. By finding and exploiting the link between two quite different theories, namely  wave interaction theory and  contour dynamics, we 
are able to  extend the analysis to the  fully non-linear domain.  

The production of Kirchhoff vortices shows that KH arising from a piecewise linear shear layer is 
very different from the classical spiraling billow type KH ensuing from a smooth shear layer. 
The characteristics of this little known KH have been investigated. The rotation and nutation of the Kirchhoff vortices
 are found to be consistent with the predictions of Kida. 
The time period of rotation of these vortices is  twice the period of nutation and the period of maximum shear layer height oscillation.  The braids connecting the Kirchhoff vortices thin exponentially fast to a length scale which is the 2D equivalent of 
 Taylor microscale. 
 
Elliptical vortical structures, similar to those found in our simulations, are  quite common in nature, especially in regions with quasi piecewise linear shear.  Examples of such vortices include 
 meddies in the Atlantic ocean, stratospheric polar vortices, and vortices in the gas giant planets like Jupiter, Neptune and Uranus. Our analysis may  
 motivate further investigation of their formation and evolution.
 

%


\end{document}